\documentclass[aps,prl,twocolumn,preprintnumbers,amsmath,amssymb,superscriptaddress,floatfix]{revtex4}
\usepackage[dvips]{graphicx}
\usepackage{mathptmx}
\usepackage{verbatim}

\begin{document}
\title{Theory of inter-edge superexchange in zigzag edge magnetism}
\author{J. Jung} \email{jeil@physics.utexas.edu}
\affiliation{Department of Physics, University of Texas at Austin, USA}
\author{T. Pereg-Barnea} 
\affiliation{Department of Physics, California Institute of Technology, Pasadena, CA 91125, USA}
\author{A. H. MacDonald} 
\affiliation{Department of Physics, University of Texas at Austin, USA}

\begin{abstract}
A graphene nanoribbon with zigzag edges has a gapped magnetic ground state with an antiferromagnetic inter-edge superexchange interaction.
We present a theory based on asymptotic properties of the Dirac-model ribbon wavefunction which predicts $W^{-2}$ and $W^{-1}$
ribbon-width dependencies for the superexchange interaction strength and the charge gap respectively.
We find that, unlike the case of conventional atomic scale superexchange, opposite spin-orientations on opposite edges of the ribbon are favored by both
kinetic and interaction energies.
\end{abstract}

\maketitle

\noindent
{\em Introduction}---
Motivated by the seminal theoretical work of Kobayashi, Fujita, Wakabayashi
and collaborators\cite{kobayashi,fujita,nakada,waka},
and by progress in graphene  
preparation\cite{Geim},
researchers have recently\cite{ezawa,brey, sasaki, hikihara, son_gap,
son_half, pisani, yazyev, joaquin,leehosik,malysheva,sasaki1}
reexamined the intriguing physics of edge magnetism in zig-zag terminated
graphene nanoribbons from a number of different points of view.
The magnetic state
is a consequence of the nearly-flat subbands which occur at the Fermi level in a neutral zig-zag ribbon, and of the orbital
character of the wavefunctions associated with these bands.
Although there is still no conclusive\cite{damagedgraphitemagnetism} experimental evidence that
a one-dimensional (1D) magnetic state occurs in ideal zigzag ribbons,
the theoretically predicted state seems likely given
that quite different electronic structure theories (from crude
Hubbard models to elaborate {\em ab initio} DFT calculations) yield consistent\cite{joaquin}
predictions and that there are at present no other ideas on how
the unusual flat bands could be accommodated in the many-electron state.
Present ribbons are far from ideal, however, and the main obstacle
to realizing this paradigmatic example of $d^{0}$ magnetism\cite{coeyd0,katzreview}
may lie in furthering recent progress\cite{han,chem_ribbon,etching}
toward chemistry and defect control at the edge.

Mean-field-theory calculations predict that the ground state of a zigzag ribbon
has unusually stiff parallel spin-alignment along each edge\cite{yazyev}
and antiferromagnetic\cite{leehosik} inter-edge superexchange
interactions.
In this Letter we present a mostly analytic theory of the ribbon width $W$ dependence
of the important inter-edge superexchange interaction.
Our theory relies on the properties of large $W$ solutions of the
continuum model approximation\cite{brey} for zigzag edges.  We predict that
interedge interactions can have a substantial influence on the properties
of these unusual 1D magnets.

\noindent
{\em Ribbon edge-state bands}---
The $\pi$-orbital tight-binding model for a finite width ribbon
yields a number of one-dimensional Bloch bands
proportional to the ribbon width \cite{malysheva}.
To a high degree of accuracy, zigzag magnetism
is a rearrangement of only\cite{joaquin} the highest occupied
($ \left| k \,\, - \right> $) and lowest unoccupied
($\left| k \,\, + \right>$) ribbon bands, whose transverse
wavefunctions are respectively odd and even functions of
carbon atom sites across the ribbon. 
Since the exchange physics which favors magnetism is local, it is revealed
most clearly by forming states in this Hilbert space which are
localized as far as possible at one edge or the other:
\begin{eqnarray}
\label{leftright}
\left| k \, L  \right>  =  \frac{1}{\sqrt{2}} \left(  \left| k \, - \right>   +    \left| k \, +   \right>    \right), \quad
\left| k \,  R \right>  =  \frac{1}{\sqrt{2}} \left(  \left| k \, + \right>     -    \left| k \, - \right>    \right)   \label{lrbasis}
\end{eqnarray}
where we have chosen the band transverse wavefunction amplitudes to be positive at the left most atom.
These L(eft), R(ight) basis states can be expanded in terms of amplitudes on atoms in the ribbon unit cell:
$\left| k \, L \right> = L_{kl} \left| k \, l \right>_B$,  $\left| k \,R \right> = R_{kl} \left| k \, l \right>_B$
with sums over sites $l$ within the unit cell implied.  It is readily verified that $L_{kl}$ and $R_{k'l'}$
are strictly locatized on opposite sublattices so that $L_{kl} R_{k'l'} = 0$ \cite{hf_edges}.
In this representation, the $\pi$-band tight-binding model Hamiltonian
$H_{TB} \left( k \right)=  t \left( k \right) \tau_{x} $
where $\tau_{x}$ is a Pauli matrix and the left-right tunneling
amplitude $t\left( k \right) = -E^{-}_{TB} \left( k \right) = E^{+}_{TB} \left( k \right)$.
The bands in the Brillouin zone $-\pi/a \le k \le \pi/a$ have periodicity $2\pi/a$
and have inversion symmetry so we can restrict our attention to the interval  $0 \le k \le \pi/a$.
Zigzag edge magnetism follows from the following tight-binding model
property \cite{brey,malysheva}. For $2\pi/3+q_e a \le \left| k \right| $,  where
$q_e = 1/ W = 2/ \sqrt{3} a N$,  the states $\left| k \, L \right>$ and $\left| k \,  R \right>$ 
are exponentially localized near their respective edges and the 
left-right hopping amplitude $t\left( k \right)$ decreases rapidly with increasing $W$. 
(Here $a = 2.46 \AA$ is the lattice constant
and $N$ is the number of atom pairs in the ribbon unit cell.)
Over this region of wavevector $|k\,L\rangle$ and $|k\,R\rangle$ are proper edge states.
\noindent

\begin{figure}[htb]
\begin{center}
\begin{tabular}{cc}
\resizebox{42.6mm}{!}{\includegraphics{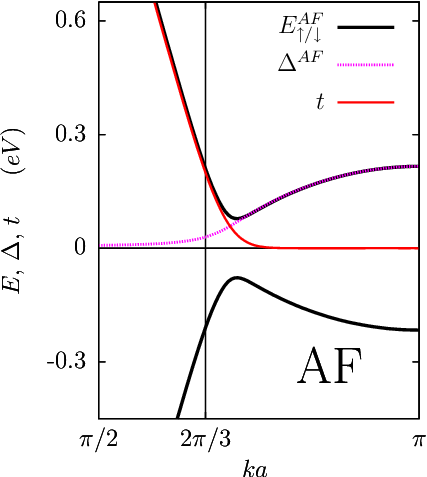}} &  \resizebox{37.4mm}{!}{\includegraphics{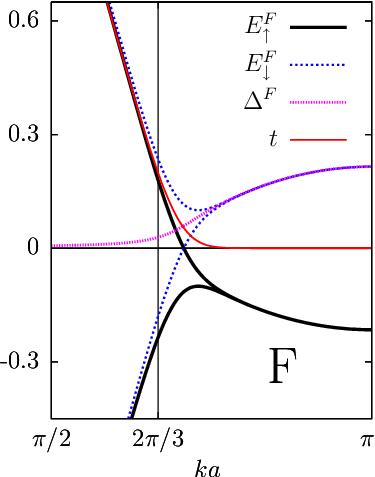}} \\
    \end{tabular}
\caption{(Color online)
Hubbard model mean-field calculations for $\gamma_0 = 2.6 eV$ and $U = 2 eV$
for a zigzag ribbon with $N=20$.
{\em Left panel}: Mean-field energy bands for the AF state:
$E^{AF} \left( k \right)= \pm \left( \Delta^{AF}\left( k \right)^{2} + t^2 \left( k \right)\right)^{1/2}$
for both spins with the low-energy states of one spin concentrated on one side of the ribbon and the
low-energy states of the opposite spin concentrated on the opposite side.
$t \left( k \right)$, the left-right hopping parameter, is quite insignificant
for this ribbon width in the edge state region.  $\Delta^{AF} \left( k \right)$
is dominant in the edge state region because of large
local spin-polarizations.
{\em Right panel}: Mean-field energy bands for the F state:
$E^{F}_{\sigma} \left( k \right) = \sigma \Delta^{F} \left( k \right) \pm \left| t \left( k \right) \right|$.
Note that $\Delta^{AF}$ and $\Delta^{F}$ are nearly identical.  The bands are periodic with periodicity $2\pi/a$ and
inversion symmetric.
}
\label{fig:hubbardmf}
\end{center}
\end{figure}

{\em Hubbard model mean-field theory}---  In order to explain our theory of the superexchange interaction
we briefly summarize the mean-field theory of the magnetic state, which is
particularly simple in the Hubbard model case.  It is instructive to contrast
two different collinear magnetic solutions of the mean-field equations, an
antiferromagnetic (AF) one in which spins have opposite orientations on opposite edges and a
ferromagnetic (F) one in which spins have the same orientation on opposite edges.
The two LR basis spin-dependent mean-field Hamiltonians are given (up to a common constant) by:
\begin{equation}
\label{hamaf}
{H}^{AF}_{\sigma} =  \left(    \begin{array}{cc}     - \sigma  \Delta^{AF}     &  t      \\
          	 t      &  \sigma \Delta^{AF}   \end{array}  \right),  \,\,
{H}^{F}_{\sigma}
	 = \left( \begin{array}{cc}   - \sigma \Delta^{F}     &  t   \\
                 t    &   - \sigma \Delta^{F}  \end{array} \right)
\end{equation}
where $\sigma = +/-$ for $\uparrow/\downarrow$ spin, the $k$-dependence of $t$ and
the self-consistent exchange potentials is implicit, and
\begin{equation}
\label{deltahb}
\Delta^{AF} \left( k \right) =
U \sum_l  L^2_{k\,\,l} \left< m_l \right>_{AF}, \quad
\Delta^{F} \left( k \right)
=  U \sum_l  L^2_{k\,\,l} \left< m_l \right>_{F}.
\end{equation}
In Eq.(~\ref{deltahb}) \, $m_l = \left(n_{l \,\uparrow} - n_{l \,\downarrow} \right) /2$
is the site-dependent spin-density
and $n_{l \sigma}$ is the spin-dependent mean occupation number at site $l$.
(Both solutions have $n_l \equiv \left(n_{l \,\uparrow} + n_{l \,\downarrow} \right) \equiv 1$, a convenient
property of neutral ribbons which can be traced to particle-hole symmetry in the paramagnetic bands.)
These self-consistent solutions are
illustrated for a $N=20$ zigzag ribbon in Fig.(~\ref{fig:hubbardmf}).  Both the F (even $m_l$) and AF (odd $m_l$) state solutions
are self-consistent.  The mean-field equations are closed by evaluating $m_l$ using
\begin{eqnarray}
\label{mmomaf}
\left< m_l \right>_{AF}&=&  \frac{a}{2 \pi}\int_{0}^{\pi/a} dk
 \left( L^{2}_{k  l} - R^{2}_{k  l}\right) \; P(k) \\
\label{mmomf}
\left< m_{l} \right>_{F}
&=&   \frac{a}{2 \pi}\int_{k_c}^{\pi/a} dk
 \left( R^{2}_{k  l} + L^{2}_{k  l}\right) \;
\end{eqnarray}
where
\begin{equation}
\label{eq:polarization}
P(k) \equiv \frac{ \Delta^{AF} \left( k \right)}{((\Delta^{AF} \left( k \right))^2 + t \left( k \right)^2)^{1/2}}
\end{equation}
is the degree of left-right edge polarization of the AF mean-field states.
In the AF case, local spin-polarization follows from the opposite left-right polarizations of
$\uparrow$ and $\downarrow$ states whereas in the $F$ case the left-right polarization vanishes and spin-polarization
follows from double occupation of $\uparrow$ bands for $k_c < |k| < \pi/a$ where
$k_c$ is the wavevector at which the $\uparrow,+$ band and the
$\downarrow,-$ band cross as illustrated in Fig. (\ref{fig:hubbardmf}).
$\Delta^{AF}$ and $\Delta^{F}$ are nearly (but not quite!) identical because their $m_l$'s differ
mostly in the middle of the ribbon which has little influence on the edge states.

\noindent
{\em Ribbon width scaling rules}---
From solutions of the graphene continuum model \cite{brey,malysheva}
we obtain for the region near $k=\pm (2\pi/3+q)$ and small $q$ the expression
\begin{eqnarray}
\label{tunneling}
t(\pm(q + 2\pi/3)) = (\sqrt{3}\gamma_0 a /2) \sqrt{q^2 - z^2},
\end{eqnarray}
where
$z$ satisfies
\begin{eqnarray}
\label{relation1}
z \,W \coth \left( z \,W \right) = q \,W, \quad \quad q > q_e \nonumber \\
z \,W/\tan \left( W z\right) = q\,W, \quad \quad q < q_e.
\end{eqnarray}
In the continuum model
\begin{equation}
R_{k l}^2 \to R_{k}^{\,2}(y) = 2 z \left( \cosh \left( 2 z y \right) -1 \right)
/ \left( \sinh\left( 2 W z \right) - 2 W z \right)
\end{equation} for $k_c < k < \pi/a $
and
\begin{equation}
R_{k}^{\,2}(y) = 4 z \sin \left( z y\right) / \left( 2 W z - \sin \left( 2 W z \right) \right).
\end{equation}
for $0 < k < k_c$. The left centered
functions can be obtained through the symmetry relation $L_{k\,l} = R_{k \, 2N-l}$.
It follows that
\begin{equation}
\label{tscaling}
 t \left( k \right) = \frac{\gamma_0}{W} \;  \tilde{t}\left( q W \right)
\end{equation}
and that
\begin{equation}
\label{rscaling}
R_{k}(y)^2 =  W^{-1}\; \chi^2(qW,y/W),
\end{equation}
where the functions $\chi^2(x)$ and $\tilde{t}(x)$ are implicitly defined by the above equations.
We have verified that these scaling relations apply accurately even in quite narrow ribbons.

\begin{figure}[htb]
\begin{center}
\begin{tabular}{cc}
\resizebox{42.2mm}{!}{\includegraphics{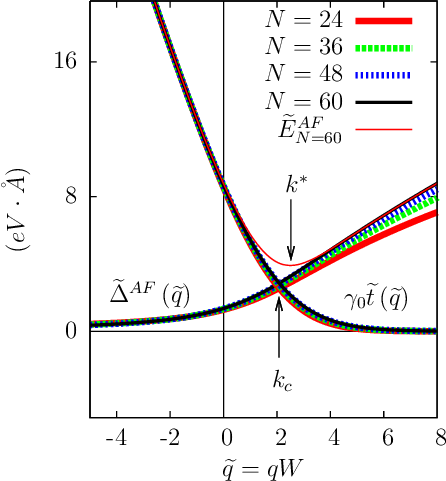}} &
\resizebox{37.8mm}{!}{\includegraphics{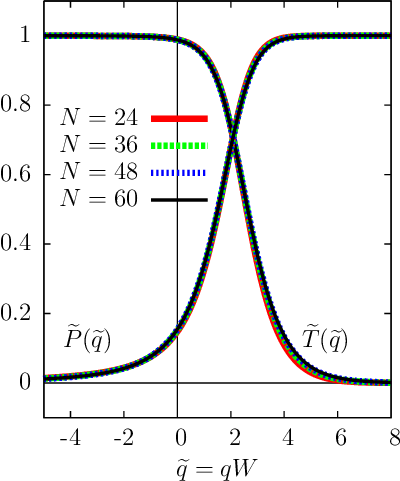}} \\
    \end{tabular}
\caption{(Color online)
{\em Left panel}: Dependence of $\widetilde{\Delta}^{AF}$ and $\gamma_0 \widetilde{t} $
on the scaled coordinate $\widetilde{q} =qW$ and the corresponding $N=60$ AF state quasiparticle bands.
Note that the self-consistently calculated $\widetilde{\Delta}^{AF}$ approaches
a well defined function at large $N$.  The positions
$k_c$ and $k^{*}$ are respectively the values of $k$ at which $\Delta^{AF} = t$ and
the band gap minimum occurs.
{\em Right Panel}: Scaling collapse of antiferromagnetic state
self-consistent left-right polarization $\widetilde{P}$ and
symmetric-antisymmetric polarization $\widetilde{T}$
represented in the scaled coordinate $\widetilde{q}$.
Note that $P^{2}+T^{2} = 1$ by definition.
}
\label{scaling}
\end{center}
\end{figure}

From Eq.(\ref{deltahb}) we see that the Hubbard model exchange potentials
depend on local spin polarizations $\left< m_l \right>$
which are large only close to the edge and approach
a well defined limit already for quite narrow ribbons; the
form of the spin-polarization near each edge is a single-edge property unrelated to
interedge interactions.  From this observation and the above scaling relations for
the zigzag edge states, we propose the following scaling rule for the form of the
exchange potential
\begin{equation}
\Delta^{AF/F} \left( k \right) = W^{-1} \; {\tilde \Delta}^{AF/F} \left(q W \right).
\label{deltascale}
\end{equation}
Since both $\widetilde{\Delta} \left( k \right)$ and $\widetilde{t} \left( k \right)$ depend only on $qW$
it follows from Eq.(~\ref{eq:polarization}) that $P(k)$ also depends only on $qW$.
We have verified numerically that this scaling relationship holds accurately 
for sufficiently wide ribbons as illustrated in Fig.(~\ref{scaling}).
\begin{figure}[htb]
\begin{center}
\begin{tabular}{c}
\resizebox{80mm}{!}{\includegraphics{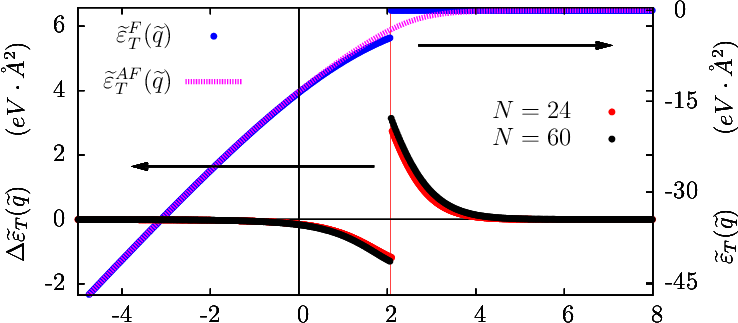}} \\
\resizebox{80mm}{!}{\includegraphics{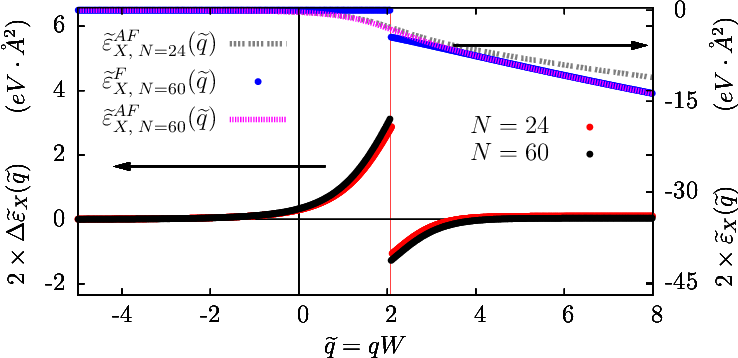}}
    \end{tabular}
\caption{(Color online)
$k$-resolved contributions to the kinetic ($\widetilde{\epsilon}_{T}$) (upper panel)
and exchange (lower panel) energies ($\widetilde{\epsilon}_{X}$) of the F and AF states
and the corresponding $F-AF$ differences as a function of the scaled
momentum coordinate $\widetilde{q} = qW$.
$\epsilon^{F}_{T}$ and $\epsilon^{AF}_{T}$ are the integrands
in the kinetic energy expression Eq.(~\ref{kin}), and
$\epsilon^{F/AF}_{X}$ is the corresponding quantity for the exchange interaction energy.
The discontinuities in the ferro case are due to the crossing between the
majority-spin symmetric and minority-spin antisymmetric bands at $k_c$,
indicated by a thin vertical line.  Although the kinetic energies
roughly double the interaction energies at most $k$ values, the exchange
contribution to superexchange is much larger because of weaker cancelation
between $|k|<k_c$ and $|k|>k_c$ regions.
}
\label{energy}
\end{center}
\end{figure}

\noindent
{\em Inter-edge interaction}---
The strength of the superexchange interaction which
determines the alignment between magnetization directions on opposite edges
is given by the total energy difference between AF and F solutions.
Because the electrostatic Hartree energies of both states are identical,
the energy difference per edge carbon atom $\Delta E$ can be separated
into band (kinetic) and exchange energy contributions:
\begin{eqnarray}
\Delta E &=& E^{F} - E^{AF} = \Delta T + \Delta E_{X}.
\end{eqnarray}
The difference of kinetic energies between AF and F solutions are determined
by contributions from occupied edge band states:
\begin{eqnarray}
\label{kin}
T^{AF} &=& - \frac{2a}{\pi} \int_0^{\pi / a} {\rm d}k \,\,   t \left( k \right)
T \left( k \right)  \quad \quad \nonumber  \\
T^{F} &=&  - \frac{2a}{\pi} \int_0^{k_c} {\rm d} k  \,\,  t \left( k \right)
\end{eqnarray}
where $T(k) = ( 1 - P^2(k) )^{1/2}$ is the symmetric-antisymmetric polarization of AF states.
The F as well as non-interacting band eigenstates have $T(k)\equiv 1$.
In the non-interacting ground state the lower energy state is fully occupied and
the total energy contains all the band energy.  Both AF and
F states sacrifice band energy contributions in the region $|k| \gtrsim 2\pi/3a$ in order to
gain interaction energy.  In the ferromagnetic case the band energy gain is sacrificed
completely for $|k|>k_c$, the wavevector at which $\Delta = t$.  At larger values of
$|k|$ both bonding and antibonding states are occupied for one spin and both are
empty for the other spin.
There is therefore an abrupt separation at $|k|=k_c$ between wavevectors which contribute to
band energy and regions which contribute to the exchange energy, discussed below.
In the AF case, on the other hand, $T$ crosses smoothly as a function of scaled wavevector $\widetilde{q}=qW$
from the kinetic energy contributing regime at small $|k|$ to the
exchange energy contributing regime at large $|k|$.  The interedge interaction
is due to this difference.  Using the scaling properties of $t$ and $\Delta$ the kinetic
energy contribution to the difference can be written as an integral over $\widetilde{q}$:
\begin{equation}
\label{kindif}
\Delta T =
\frac{2a \gamma_0 }{\pi W^2 } \left[
\int_{-\infty}^{\widetilde{q}_c} \; {\rm d}\widetilde{q} \; \,  \widetilde{t}(\widetilde{q}) \;
[\widetilde{T}(\widetilde{q})-1]
+  \int_{\widetilde{q}_c}^{\infty} \; {\rm d}\widetilde{q} \; \, \widetilde{t}(\widetilde{q}) \;
\widetilde{T}(\widetilde{q})\right].
\end{equation}
The integrals in Eq.(~\ref{kindif}) converge at $-\infty$ because $\widetilde{T}(\widetilde{q})$
approaches $1$ rapidly and at $\infty$ because $\widetilde{T}(\widetilde{q})$ approaches $0$ rapidly.
The contribution from $\widetilde{q} < \widetilde{q}_c$ is negative while the contribution from
$\widetilde{q} > \widetilde{q}_c$ is
positive.  Substantial cancelation leads to a small kinetic energy contribution to $\Delta E$.

\begin{figure}[htb]
\begin{center}
\begin{tabular}{cc}
\resizebox{39.85mm}{!}{\includegraphics{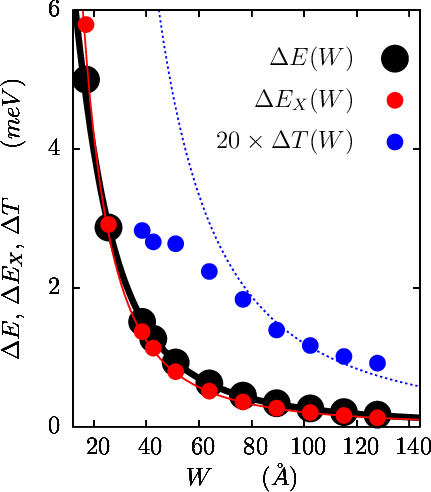}} &
\resizebox{38.15mm}{!}{\includegraphics{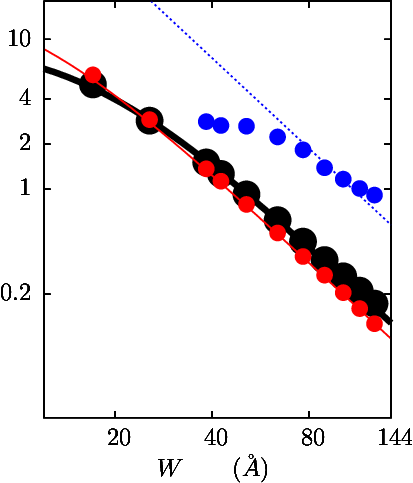}} \\
    \end{tabular}
\caption{(Color online) 
The difference between the F and AF states in total energy ($\Delta E$), exchange energy ($\Delta E_X$) and kinetic energy ($\Delta T$) plotted on a linear scale (left) and a logarithmic scale (right).
The total energy difference follows a
$W^{-2}$ decay law at large $W$ and is dominated by exchange energy contribution.
The kinetic energy contribution is substantially smaller and
the asymptotic decay law develops only for sufficiently large ribbon width.
The total energy $\Delta E $ was fitted with
$2.7/(W^{2} + 280)$, the exchange energy $\Delta E_X$
with $2.1/(W^{2} + 100)$ and the fitting for the
kinetic contribution was obtained
from the difference between both resulting in $0.6/W^2$
(represented with $\times 20$ magnification),
all terms given in $eV \AA^{-2}$ units.
We used a 12K $k$-point sampling density in the 1D Brillouin zone.
}
\label{fig4}
\end{center}
\end{figure}

The exchange energy integrands satisfy the scaling relations
similar to the kinetic terms and we can
write the exchange energy difference as
\begin{widetext}
\begin{equation}
\label{exdif}
\Delta E_{X} =
\frac{a}{\pi W^2} \Big[
\int_{-\infty}^{\widetilde{q}_c} \; {\rm d}\widetilde{q} \; \,  \widetilde{\Delta}^{AF}(\widetilde{q}) \;
\widetilde{P}(\widetilde{q})
+  \int_{\widetilde{q}_c}^{\infty} \; {\rm d}\widetilde{q} \; \, \widetilde{\Delta}^{AF}(\widetilde{q}) \;
[\widetilde{P}(\widetilde{q})-1] \Big]
+ \frac{1}{3} \;  (\Delta^{AF}(\pi/a)-\Delta^{F}(\pi/a))
\end{equation}
The first two terms are similar to the corresponding band energy contributions, with
the discontinuity at $q_c$ again due to the band crossing in the ferromagnetic state.
We write this contribution to the superexchange interaction as $J_{X}/W^2$.
An additional contribution appears because $\Delta^{AF}$ and
$\Delta^{F}$ are not quite identical for $\widetilde{q} \to \infty$.
\end{widetext}
In the Hubbard model we can relate the asymptotic difference in $\Delta$ to the difference in
spin polarization on the edge atom:
$  \delta \Delta \equiv \Delta^{AF} \left( \pi / a \right) - \Delta^{F} \left( \pi / a \right)
 =  U \left( \left< m_{edge} \right>_{AF} - \left< m_{edge} \right>_{F} \right)$.
Labeling the leftmost site as site 1, noting that $R^{2}_{k \, 1} = 0$ and
 recalling the definitions of
 $\left< m_l \right> $ in Eqs. (\ref{mmomaf},\ref{mmomf}) we find that
 \begin{eqnarray}
 \delta \Delta &=&
 \frac{a}{2\pi}  \int_{0}^{k_c} {\rm d}k
  \,\, L_{k\, 1}^{2}  P \left( k \right)
+  \frac{a}{2\pi}  \int_{k_c}^{\pi/a}   {\rm d} k
 \,\, L^2_{k \, 1} \, \left( 1 - P \left( k \right) \right)    \nonumber \\
    &\equiv &  \frac{3J_{\delta m}}{W^2}    \quad \quad
\label{jdelta}
\end{eqnarray}
Adding the three contributions, the total superexchange interaction is
\begin{eqnarray}
\Delta E \left( W \right) = W^{-2} \left( J_{K} + J_{X} + J_{\delta m} \right).
\end{eqnarray}
For $\gamma_0 = 2.6 eV$ and on site repulsion $U = 2.0 eV$ that results in band
gaps similar to LDA \cite{joaquin},
we find that the kinetic and exchange contributions to the interaction are $J_{K} = 0.6 eV \cdot \AA^{2}$,
and $J_{X} + J_{\delta m} = 2.1 eV\cdot \AA^2$. (See Fig.(~\ref{fig4}).)  Separately
$J_{\delta m} \simeq 1.15  eV \cdot \AA^2$ implying that the interaction contribution is composed
in approximately equal measures
of contributions from $q$ near $q_c$ and
contributions far in the edge regime.

\noindent
{\em Conclusions and discussions}---  Our analysis shows that the antiferromagnetic
inter-edge superexchange interaction in magnetic zigzag nanoribbons is the sum of three
contributions (band-energy, exchange-energy, and edge spin-polarization) all of which
arise from a region of the ribbons's 1D Brillouin-zone which is centered on
$|k|=2\pi/3a$ and scales in width as $1/W$.  Unlike the familiar case of atomic-scale
superexchange interactions, in which antiferromagnetic spin-arrangements lower
the kinetic energy at a cost in interaction energy, all three contributions have the
same sign - with the kinetic contribution being substantially smaller in magnitude.
Our conclusions rest primarily on analytic properties of continuum model solutions to the
$\pi$-band model for zig-zag nanoribbons and depend on the particle-hole symmetry
of graphene's conduction and valence bands.  We have demonstrated numerically that the continuum model predictions are
accurate, even in narrow nanoribbons.  Although some details of our analysis depend on
the simplified Hubbard model we use, we expect that the scaling properties of
states near $|k|=2\pi/3a$ to be general and that our qualitative
conclusions will apply to any mean-field-theory treatment of zigzag ribbon magnetism.

Collective spin-behavior is expected\cite{yazyev} to be important in zigzag ribbon magnets,
even though they are one-dimensional, because of the exceptionally strong
exchange interactions along each edge.  Assuming that the magnetic anisotropy
(which is expected to be weak) is important only at low-temperatures, the correlation length $\xi$
along an isolated zigzag edge is estimated\cite{yazyev} to be $\sim 3000 \AA/T[K]$.
Since the interedge interaction arises from an interval of $k$-space with width $1/W$, its
range along the edge will be $\sim W$.  When $\xi$ is smaller than $W$, interedge
interactions will have little influence.  For $\xi$ larger than $W$, the interedge
interactions will help suppress thermal magnetization fluctuations.

{\em Acknowledgment}--- The authors gratefully acknowledge
helpful interactions with Rafi Bistritzer, Jason Hill, Hsiu-Hau Lin,
Nikolai Sinitsyn, Joaqu\'{\i}n Fern\'{a}ndez-Rossier, Juan Jos\'{e} Palacios 
and Cheng-Ching Wang.  This work was supported by
the Welch Foundation, NRI-SWAN, ARO, DOE
and by the Spanish Ministry of Education through the MEC-Fulbright program.

\end{document}